\documentclass[aps,prb,twocolumn]{revtex4}
\usepackage{graphicx}
\usepackage{amssymb}
\usepackage{amsmath}
\usepackage{bm}
\usepackage{color}

\begin{document}
\title{Dynamics and stability of dark solitons in
       exciton-polariton condensates}
       
\author{Lev A. Smirnov$^{1,2,3}$}
\author{Daria A. Smirnova$^1$}
\author{Elena A. Ostrovskaya$^1$}
\author{Yuri S. Kivshar$^1$}

\affiliation{$^1$Nonlinear Physics Centre, Research School of Physics and Engineering, The Australian National University, Canberra ACT 0200, Australia \\
$^2$Institute of Applied Physics, Russian Academy of Sciences, Nizhny Novgorod 603950, Russia \\
$^3$University of Nizhny Novgorod, Nizhny Novgorod 603950, Russia}

\begin{abstract}
We present a comprehensive analytical theory of localized nonlinear excitations -- dark solitons, supported by an incoherently pumped, spatially homogeneous exciton-polariton condensate. We show that, in contrast to dark solitons in conservative systems, these nonlinear excitations "relax" by blending with the background at a finite time, which critically depends on the parameters of the condensate. Our analytical results for trajectory and lifetime are in excellent agreement with direct numerical simulations of the open-dissipative mean-field model. In addition, we show that transverse instability of quasi-one-dimensional dark stripes in a two-dimensional open-dissipative condensate demonstrates features that are entirely absent in conservative systems, as creation of vortex-antivortex pairs  competes with the soliton relaxation process.
\end{abstract}

\maketitle
\section{Introduction}\label{sec:Introduction} 
Bose-Einstein condensates of (exciton-)polaritons created in semiconductor microcavities in a strong light-matter interaction regime are in the focus of exciting new research area, where quantum and non-equilibrium  properties can be studied within the same physical system \cite{Yamamoto_RMP,RevModPhys_2013}. The non-equilibrium, open-dissipative behaviour is intrinsic to this system since polaritons are subject to rapid radiative decay, and their population is maintained due to an optical pumping. On the other hand, the ability of these bosoinc quasi-particles to condense into a macroscopically occupied quantum coherent state has prompted vigorous investigations of analogies between the polariton condensate and intrinsically equilibrium, conservative condensates of atomic gases which exist under conditions of careful isolation from the environment \cite{PitaevskiiStringari2003,PethickSmith2008,Kevrekidis2008}. Many features of atomic BEC have been successfully demonstrated in polariton condensates, such as long-range coherence \cite{pBEC_coherence_PRL_09,pBEC_coherence_PNAS_12}, superfluid flow \cite{Amo_superfluidity,Amo_fluid_dynamics}, and quantized vortices \cite{Vortex_pairs_Amo,Vortex_pairs_Deveaud,Roumpos_vortex}. 

\begin{figure}[b]
\includegraphics[width=8.5cm]{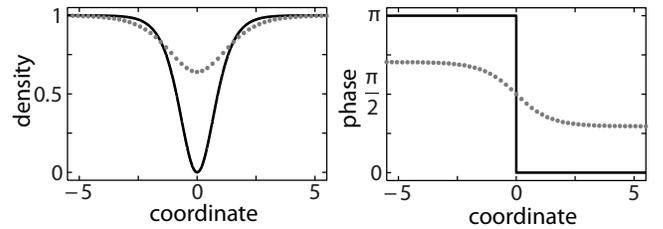}
\caption{\label{fig:intro}  Schematics of the condensate density, $|\Psi|^2$, and phase of the condensate order parameter (macrosopic wavefunction), ${\rm arg}(\Psi)$, coresponding to a stationary one-dimensional dark soliton (solid) and a moving "gray" soliton (dotted line).}
\end{figure}

Nonlinearity of the polariton condensate is inherited from strongly and repulsively interacting excitons. This results in the nonlinear behaviour akin to that of matter waves (atomic condensates) with a positive scattering length or optical waves in a nonlinear defocusing media. Consequently, the most basic form of a {\em nonlinear collective excitation} in this system is a dark soliton characterised by a density dip and an associated phase gradient (see Fig. \ref{fig:intro}). So far, dark solitons and their dynamics have been observed \cite{DS_Amo_Science_11,DSInstability_PolaritonBEC_2011,DSInstability_PolaritonBEC_2012,pBEC_ring13} and analysed \cite{DS_theory_08,DS_obstacle_theory} mostly in the polariton condensates coherently and resonantly driven by a pumping laser. Formation and behaviour of dark solitons in an {\em incoherently pumped} polariton condesate with a spontaneously  established coherence, is much less understood. Several numerical studies \cite{1DpBECth2011,DSinc2013,train_pBEC} suggest that analogies between dark solitons in polariton condensates and those in matter or optical waves \cite{Kivshar_1998,Analogies_2004,Frantzeskakis_2010} cannot be taken too far. The open-dissipative nature of polariton condensates lends unique, and so far little explored, features to localized nonlinear excitations.

In this paper, we analyse, both analytically and numerically, dynamics and stability of dark solitons in an incoherently excited, spatially homogeneous polariton condensate.  We obtain a number of analytical results that elucidate the influence of open-dissipative nature of the system on the properties of localized nonlinear excitations. In particular, we show that dark soliton-like structures supported by a modulationally stable background have a finite lifetime, and do not remain spatially localized. We also show that the hallmark of quasi-one-dimensional dark solitons in two- (or three-) dimensional optical and matter waves, the so-called {\em snake instability} leading to formation of vortex pairs \cite{Tikhonenko_1996,PhysRevLett.86.2926_2001}, can be completely inhibited in a two-dimensional open-dissipative condensate.

The paper is organised as follows. First, we consider the mean-field model of a polariton condensate subject to incoherent off-resonant optical excitation, discuss the homogeneous steady state and introduce appropriate scalings. In Sec. \ref{BdG}, we revisit the modulational stability analysis of the homogeneous state by means of Bogoliubov-de Gennes approach, and derive simple analytical expressions for the boundaries of unstable regions in the parameter space. Next, in Sec. \ref{sec:DSDynamic}, we construct general asymptotic theory for the dynamics of dark solitonic excitations supported by  a homogeneous condensate. By comparison with numerical simulations, we demonstrate validity of the analytical theory in the regimes of weak and strong pumping. Section \ref{sec:DSInstability} focuses on transverse instability of quasi-one-dimensional dark solitons in a two-dimensional condensate and on the effect of competing time scales between the processes of soliton relaxation and instability development. Finally, we conclude with a brief summary of our results.

\section{The model}\label{sec:Model} 
Within a mean-field description, the macroscopic wave function, $\Psi\!\left(\vec{r},t\right)$, of an incoherently, far off-resonantly pumped polariton condensate is governed by a generalized open-dissipative Gross-Pitaevskii (GP) equation coupled to a rate equation for a density, $n_{R}\!\left(\vec{r},t\right)$, of an uncondensed reservoir of high-energy near-excitonic polaritons~\cite{WoutersCarusotto_2007,RevModPhys_2013}: 
\begin{eqnarray}
\label{eq:GPeq1}
i\hbar\frac{\partial\Psi}{\partial t}&=&\left[-\frac{\hbar^{2}}{2M}\Delta+U\left(\vec{r},t\right)+\frac{i\hbar}{2}\left(R n_{R}-\gamma_{C}\right)\right]\Psi, \\ 
\label{eq:nReq1}
\frac{\partial n_{R}}{\partial t}&=&-\left(\gamma_{R}+R\left|\Psi\right|^{2}\right)n_{R}+P\left(\vec{r},t\right),
\end{eqnarray}
where $U\left(\vec{r},t\right)=g_{C}\left|\Psi\right|^{2}+V_{\rm ext}+g_{R}n_{R}$ is an effective potential, combining blueshifts due to polariton condensate interactions and polariton-reservoir interactions. Here, $g_C$ is the strength of nonlinear interaction of polaritons,  $g_{R}$ is the condensate coupling to the reservoir, $R$ stands for the stimulated scattering rate, and $\gamma_{C}$ is the rate of loss of condensate polaritons. High energy, exciton-like polaritons are injected into the reservoir by laser pump $P\!\left(\vec{r},t\right)$ and relax at the reservoir loss rate $\gamma_{R}$.

We recall that under continuous-wave ({\em cw}) and spatially uniform pumping, $P\left(\vec{r},t\right)=P_{0}={\rm const}$, the steady state solution is sought in the form~\cite{WoutersCarusotto_2007,RevModPhys_2013}:
\begin{equation}
\label{eq:hss0}
\Psi\left(\vec{r},t\right)=\Psi_{0}=\sqrt{n^0_C}e^{-i(E_{0}/\hbar)t}, \,\, 
n_{R}\left(\vec{r},t\right)=n^0_{R},
\end{equation}
where quantities $n^0_C$ and $n^0_R$ are constant. For a weak pump $P_{0}$, the condensate is absent $n^0_C=0$, while the reservoir density is proportional to the pump intensity $n^0_R=P_{0}/\gamma_R$. Exact balance of loss and gain is achieved at the threshold value $P_{th}=\gamma_{R}\gamma_C/R$. Above this threshold, when $P_{0}>P_{th}$, the solution~(\ref{eq:hss0}) with $n^0_C=0$ becomes unstable and the condensate appears. The steady homogeneous condensate and reservoir densities are expressed as follows:
\begin{equation}
\label{eq:hss1}
n^0_C=\left(P_0-P_{ th}\right)/\gamma_C, \quad
n^0_R=n^{th}_{R}=\gamma_C/R,
\end{equation}
and the condensate energy is
\begin{equation}
\label{eq:omegahss}
E_0=g_C n^0_C+g_R n^{th}_R.
\end{equation}\par
The model~(\ref{eq:GPeq1}),~(\ref{eq:nReq1}) can be rewritten in a dimensionless form by using the scaling unit of healing length $r_h=\hbar/(M c_s)$ and time $\tau_{0}=r_h/c_s$, where $c_s=(g_C n^*_C/M)^{1/2}$ is a local sound velocity in the condensate, and $n^*_C$ is a characteristic value of the condensate density. The dimensionless equations for the normalized condensate wavefunction ${\bar \Psi}=\Psi(n^*_C)^{-1/2}$ and reservoir density ${\bar n}_R=n_R/n^*_C$ take the form:
\begin{eqnarray}
\label{eq:GPeq2}
i\frac{\partial\bar{\Psi}}{\partial t}&=&\left[-\frac{1}{2}\Delta+{\bar U}+\frac{i}{2}(\bar{R}\bar{n}_{R}-\bar{\gamma}_{C})\right]\bar{\Psi}, \\ 
\label{eq:nReq2}
\frac{\partial\bar{n}_{R}}{\partial t}&=&-(\bar{\gamma}_{R}+\bar{R}\left|\bar{\Psi}\right|^{2})\bar{n}_{R}+\bar{P}.
\end{eqnarray}
where ${\bar U}=U/(g_C n^*_C)$,  ${\bar P}=P/(g_Cn^{*2}_C)$, and we omitted the bars over the dimensionless time variable. The corresponding dimensionless parameters are:
\begin{equation}
\bar{\textsl{g}}_{R}=\frac{\textsl{g}_{R}}{\textsl{g}_{C}},\,\,
\bar{\gamma}_{C}=\frac{\hbar\gamma_{C}}{\textsl{g}_{C}{n^*_{C}}}=\frac{\gamma_{C}}{\gamma_{R}}\bar{\gamma}_{R},\,\,
\bar{R}=\frac{\hbar R}{\textsl{g}_{C}}.
\end{equation}\par
For a {\em cw} background it is convenient to choose $n^*_C\equiv n^0_C$. Thus the homogeneous steady state~(\ref{eq:hss0}),~(\ref{eq:hss1}) and~(\ref{eq:omegahss}) can be rewritten as:
\begin{equation}
\label{eq:hss2}
\bar{\Psi}_{0}=\exp\left(-i\bar{\omega}_{0}t\right),\,\,
\bar{\omega}_{0}=1+\bar{g}_{R}\bar{n}^{th}_R,\,\,
\bar{n}^{\rm th}_R=\bar{\gamma}_{C}/\bar{R}.
\end{equation}
Our aim for the rest of this work is to construct a theory for propagation and stability of {\em nonlinear waves} -- finite amplitude collective excitations 
(such as dark solitons, quantum vortices, etc) of a homogeneous condensate under incoherent uniform pumping. To this end, we consider perturbations of the condensate wave function and the reservoir density in the following general form:
\begin{equation}
\bar{\Psi}=\bar{\Psi}_{0}\left(t\right)\psi \left(\vec{r},t\right); \qquad \bar{n}_{R}=\bar{n}^{th}_{R}+m_{R}\left(\vec{r},t\right).
\end{equation}
The perturbations $\psi\!\left(\vec{r},t\right)$ and $m_{R}\!\left(\vec{r},t\right)$ are governed by the dynamical equations:
\begin{eqnarray}
\label{eq:psieq}
i\frac{\partial\psi}{\partial t}=\left[-\frac{1}{2}\Delta-(1-\left|\psi\right|^{2})+\bar{g}_{R}m_{R}+\frac{i}{2}\bar{R}m_{R}\right]\psi,\\
\label{eq:mReq}
\frac{\partial m_{R}}{\partial t}=\bar{\gamma}_{C}\Bigl(1-\left|\psi\right|^{2}\Bigr)-\bar{\gamma}_{R}m_{R}-\bar{R}\left|\psi\right|^{2}m_{R},
\end{eqnarray}
where $\bar{\gamma}_{R}=\bar{R}/\left(P_{0}/P_{th}-1\right)$.

\section{Stability of a homogeneous condensate}\label{BdG}
First of all, it is essential to analyze stability of the homogeneous background with respect to weak perturbations~\cite{RevModPhys_2013, WoutersCarusotto_2007, NegativeBogoliubovDispersion_2013}. Here we revisit the already known results on spectra of elementary excitations for the open-dissipative model and derive new analytical criteria for the stability of the homogeneous background.

The system of equations~(\ref{eq:psieq}),~(\ref{eq:mReq}) has a solution $\left|\psi\right|^{2}\!=\!1$, $m_{R}\!=\!0$, corresponding to a homogeneous distribution of the condensate density $n_{C}$ and the polariton reservoir density $n_{R}$. The linear stability of this stationary state can be analyzed by means of Bogoliubov-de Gennes approach~\cite{PitaevskiiStringari2003, PethickSmith2008} by introducing small perturbations of the homogeneous background in the form~\cite{WoutersCarusotto_2007}:
\begin{eqnarray}
\label{eq:BGrepresentation:psi}
\psi=1+\epsilon\sum_{\vec{k}}\left[a_{\vec{k}}e^{\left (-i\omega t+i\vec{k}\vec{r} \right )}+b_{\vec{k}}e^{ \left (i \omega^*t-i\vec{k}\vec{r} \right )} \right],\\
\label{eq:BGrepresentation:mR}
m_{R}=\epsilon\sum_{\vec{k}}\left[c_{\vec{k}}e^{\left(-i\omega t+i\vec{k}\vec{r}\right)}+c^*_{\vec{k}}e^{\left(i\omega^*t-i\vec{k}\vec{r}\right)}\right],
\end{eqnarray}
where $\epsilon\!\ll\!1$ is a small parameter. Substituting~(\ref{eq:BGrepresentation:psi}),~(\ref{eq:BGrepresentation:mR}) into~(\ref{eq:psieq}),~(\ref{eq:mReq}) and keeping terms linear in $\epsilon$, we obtain the eigenvalue problem for elementary excitations~\cite{WoutersCarusotto_2007}.
%
Solution of the eigenvalue problem yields the dispersion relation, which is a cubic equation with respect to $\omega(k)$
\begin{equation}
\omega^{3}+i\!\left(\bar{\gamma}_{R}+\bar{R}\right)\omega^{2}-\left(\omega_{B}^{2}+\bar{R}\bar{\gamma}_{C}\right)\omega=f(k),
\label{eq:DispersionRelation}
\end{equation}
where $f(k)=i\left(\bar{\gamma}_{R}-\bar{R}\right)\omega_{B}^{2}-i\bar{g}_{R}\bar{\gamma}_{C}k^{2}$, and $\omega^2_{B}\left(k\right)=k^{2}+{k^{4}}/4$ is the standard Bogoliubov dispersion relation for an equilibrium (atomic) condensate~\cite{PitaevskiiStringari2003, PethickSmith2008, NegativeBogoliubovDispersion_2013}.

For a fixed $k$, Eq. (\ref{eq:DispersionRelation}) has three roots $\omega_{i}$, $j=1,2,3$, determining eigenfrequencies of elementary excitations. According to the Cardano formula, depending on the values of parameters $\bar{g}_{R}$, $\bar{\gamma}_{C}$, $\bar{\gamma}_{R}$, $\bar{R}$, either all three roots are purely imaginary, or one root $\omega_{1}$ is purely imaginary, while $\omega_{2}$ and $\omega_{3}$ have real parts of the opposite signs.

Accordingly, for fixed values of the system parameters, the continuous functions $\omega_{j}(k)$ represent three dispersion branches of the linear perturbations. When $\bar{\textsl{g}}_{R}$, $\bar{\gamma}_{C}$, $\bar{\gamma}_{R}$ and $\bar{R}$ are vanishingly small, i.e. when the influence of the reservoir on the condensate dynamics is eliminated, two roots $\omega_{2}\!\left(k\right)$ and $\omega_{3}\!\left(k\right)$ correspond to Bogoliubov dispersion law $\pm\omega_{B}\!\left(k\right)$ of the equilibrium condensate, whereas $\omega_{1}\!\left(k\right)$ characterizes aperiodic behaviour of reservoir perturbations~\cite{RevModPhys_2013, WoutersCarusotto_2007}.

In general, the roots of Eq.~(\ref{eq:DispersionRelation}) are complex: $\omega_{j}\!\left(k\right)\!=\!\Omega_{j}+i\Gamma_{j}$. If $\Gamma_{j}\!\left(k\right)\!>\!0$, the homogeneous condensate density distribution is {\em modulationally (dynamically) unstable}, since its density modulations grow in time exponentially, while the opposite case of decaying perturbations, $\Gamma_{j}\!\left(k\right)<0$, corresponds to a stable regime.

In order for a {\em cw} background to be stable, perturbations should not grow for all values of $k$, i.e. the following condition should be satisfied: $
\Gamma_{j}\!\left(k\right)\leq0$ for all $j$. This condition is {\em always fulfilled} for $k\!=\!0$ and $k\!\rightarrow\!\infty$. 
Indeed, at $k=0$, one of the roots of Eq.~(\ref{eq:DispersionRelation}) is zero, and the other two roots have negative imaginary parts. The former is associated with a Goldstone mode that can be understood as a slow rotation of the condensate phase~\cite{RevModPhys_2013, WoutersCarusotto_2007}. This mode exhibits diffusive properties at small values of $k$~\cite{RevModPhys_2013, WoutersCarusotto_2007}.
 In the opposite limit, $\left(k\!\rightarrow\!\infty\right)$, one can show that all three branches $\omega_{j}\!\left(k\right)$ have negative imaginary parts: $\omega_{1}\!\left(k\right)\approx-i\left(\bar{\gamma}_{R}+\bar{R}\right)$,
$\omega_{2,3}\!\left(k\right)\approx\pm\omega_{B}\!\left(k\right)-\bigl.i\bar{\gamma}_{C}\bar{R}k^{2}\bigr/2\omega_{B}^{2}\!\left(k\right)$. Thus all perturbations decay in time.

If $\Gamma_{j}\!\left(k\right)>0$ in some range of $k$, two values $k\!=\! k_{1,2}^{(j\ast)}$ correspond to the boundaries of the steady-state instability domain, appearing at least on one of the dispersion branches $\omega_{j}\!\left(k\right)$, when $\Gamma_{j}\!\left(k_{1,2}^{(j\ast)}\right)\!=\!0$. The critical values $k\!=\! k_{1,2}^{(j\ast)}$, separating stable and unstable regions, can be found from Eq.~(\ref{eq:DispersionRelation}) by substituting the eigenfrequencies with $\Gamma=0$, i.e. $\omega\!\equiv\!\Omega$. As a result, we conclude that, for $\Gamma_j$ to change sign, two equalities should be satisfied simultaneously:
\begin{eqnarray}
\Omega \left(\Omega^{2}-\omega_{B}^{2}-\bar{R}\bar{\gamma}_{C}\right)=0,\\
\left(\bar{\gamma}_{R}+\bar{R}\right)\Omega^{2}-\left(\bar{\gamma}{}_{R}+\bar{R}\right)\!\omega_{B}^{2}+\bar{g}_{R}\bar{\gamma}_{C}k^{2}=0.
\end{eqnarray}
Since all the parameters of our physical system are positive and real, this is possible only if $\Omega\!=\!0$, i.{\,}e. for a purely imaginary dispersion branch $\omega_{1}\!\left(k\right)$. Furthermore, $k_{1}^{(1\ast)}\!\equiv\!0$, and the other critical value
\begin{equation}
k_{2}^{(1\ast)}=2\sqrt{\frac{\bar{g}_R\bar{\gamma}_C}{\bar{\gamma}_R+\bar{R}}-1}
\end{equation}
exists only under condition
\begin{equation}
\bar{\textsl{g}}_{R}\bar{\gamma}_{C}>\left(\bar{\gamma}{}_{R}+\bar{R}\right)\label{eq:InstabilityCondition}.
\end{equation}
Thus, if the condition~(\ref{eq:InstabilityCondition}) is fulfilled, the perturbation with wavenumber $k$ from the interval $[k_{1}^{(1\ast)},k_{2}^{(1\ast)}]$ grow and a homogeneous condensate is {\em modulationally unstable}. 
Hereafter, we will be interested in the opposite case of the modulationally stable {\em cw} condensate. In our original variables the {\em stability} condition for the {\em cw} background takes the form:
\begin{equation}
\label{eq:FullStabilityCondition_DimensionVariables}
\frac{P_{0}}{P_{th}}>\frac{g_{R}}{g_{C}}\frac{\gamma_{C}}{\gamma_{R}}.
\end{equation}
Within the framework of our open-dissipative model the inequality~(\ref{eq:FullStabilityCondition_DimensionVariables}) means that a homogeneous steady state of  a polariton condensate is stable for {\em all} values of the pump intensity $P_{0}>P_{th}$ only under condition that 
\begin{equation}
\gamma_{R}\bigl/\gamma_{C}\bigr.>\textsl{g}_{R}\bigl/\textsl{g}_{C}.
\end{equation}
If this relation is violated, there is a range $P_{th}<P_{0}<\left(g_{R}\gamma_{C}/g_{C}\gamma_{R}\right)P_{th}$, where a homogeneous background is modulationally unstable. 

In what follows, we restrict our consideration to the dynamics of nonlinear waves propagating in a modulationally stable polariton condensate.
\section{Dynamics of one-dimensional dark solitons}\label{sec:DSDynamic}

The simplest nonlinear excitation supported by a spatially homogeneous modulationaly stable condensate with repulsive inter-particle interactions is a one-dimensional (1D) dark soliton -- a localized dip in the condensate density with an associated phase gradient (Fig. \ref{fig:intro}). Such structures may exist in a condensate with a reduced dimensionality, e.g., that contained in a microwire. Soliton "stripes" in a two-dimensional (2D) condensate, spatially uniform along one of the dimensions, can also be treated as quasi-one-dimensional structures. Strictly speaking, as we show below, the main features of a soliton, e.g. propagation through the supporting media without any change in shape or velocity are {\em absent} in open-dissipative condensate. The solitonic nature of the nonlinear excitations, such as their spatial localization, may be maintained over a period of time determined by the system parameters, however the intrinsic dissipation causes the dark states to delocalize and blend with the background. Remarkably, within the framework of our model, we can derive simple analytical expressions for the velocity and lifetime of the dark solitonic excitations in polariton condensates.

\subsection{General asymptotic description}\label{subsec:AsymptoticDescription}
If the perturbation $m_{R}\!\left(\vec{r},t\right)$ of the reservoir density is set to zero, Eq.~(\ref{eq:psieq}) becomes a nonlinear Schr\"{o}dinger~(NLS) equation with a "defocusing" nonlinearity:
\begin{equation}
\label{eq:NLS}
i\frac{\partial\psi}{\partial t}+\frac{1}{2}\Delta\psi+\Bigl(1-\left|\psi\right|^{2}\Bigr)\psi=0
\end{equation}
It is well known that this equation has a single-parameter family of solutions in the form of one-dimensional dark solitons moving (for definiteness, along the $x$ axis) at constant velocity $v_{s}$ $\left(0\!\leq\!\left|v_{s}\right|\!<\!1\right)$~\cite{Kivshar_1998, Frantzeskakis_2010, Kevrekidis-art-2008, Kevrekidis2008}. In the moving reference frame, $\xi=\left(x-v_{s}t\right)$, a dark soliton is described by a wave function of the form:
\begin{equation}
\label{eq:ds}
\psi_{s}\left(\xi,v_{s}\right)=\sqrt{1-v_{s}^{2}}\tanh\Bigl(\sqrt{1-v_{s}^{2}}{\,}\xi\Bigr)+iv_{s},
\end{equation}
which satisfies the stationary equation:
\begin{equation}
\label{eq:dsNLS}
-iv_{s}\frac{\partial\psi_{s}}{\partial\xi}+\frac{1}{2}\frac{\partial^{2}\psi_{s}}{\partial\xi^{2}}+\left(1-\left|\psi_{s}\right|^{2}\right)\psi_{s}=0
\end{equation}
and the boundary conditions at infinity
\begin{equation}
\label{eq:bouncond}
\psi_{s}\!\left(\sqrt{\xi^{2}+y^{2}}\rightarrow\infty\right)\rightarrow 1.
\end{equation}
A stationary soliton ($v_{s}= 0$) is often referred to as a "black" soliton, where the BEC density drops to zero, and the phase of the wavefunction $\psi_{s}\left(\xi\right)$ has a $\pi$ phase jump across the soliton profile (see Fig. \ref{fig:intro}). In a moving ("gray") soliton, the minimum value of density $n^{\rm min}_{C}=\left|\psi_{s}\left(\xi=0\right)\right|^{2}$ increases in proportion to the square of the soliton velocity $n^{\rm min}_{C}=v_{s}^{2}$, while the phase changes smoothly. When $v_{s} \rightarrow 1$, the minimum density attains the background value, i.e. $n^{\rm min}_{C}\to 1$ in our dimensionless units. Simultaneously, the soliton width tends to infinity: $\Lambda_{s}=(1-v_{s}^{2})^{-1/2}\to \infty$.

Next, we consider the limit $m_{R}\left(\vec{r},t\right) \ll 1$, which, as will be shown below, is quite realistic. The right-hand-side terms of Eq.~(\ref{eq:psieq}) proportional to $m_R$,
\begin{equation}
\label{eq:RightHandSide}
\mathcal{R}\!\left(m_{R},\psi\right)=\left(\bar{\textsl{g}}_{R}+\frac{i}{2}\bar{R}\right)\!m_{R}\psi,
\end{equation}
can be treated as small perturbations. A polariton condensate with a weak coupling to the reservoir should support solitonic structures similar to~(\ref{eq:ds}), but with a slowly varying in time velocity $v_{s}=v_{s}\left(\mu t\right)$. Here we introduced a small parameter $\mu\ll1$, which will be defined differently for each of the cases considered below. Such solutions can be sought in the form of asymptotic expansion in $\mu$:
\begin{eqnarray}
\label{eq:AsymptoticSeries}
\psi\!\left(\vec{r},t\right)=\psi_{s}\!\left(\xi,v_{s}\!\left(\mu t\right)\right)+\sum_{j=1}^{\infty}\mu^{j}\psi_{j}\!\left(\xi,\mu t\right),\\
m_{R}\!\left(\vec{r},t\right)=\mu m^0_{R}\left(\xi,\mu t\right)+\sum_{j=1}^{\infty}\mu^{\left(j+1\right)}{m_{R}}_{j}\!\left(\xi,\mu t\right).
\end{eqnarray}
Basic understanding of perturbation-induced dynamics for a dark soliton can be deduced from analysis of evolution equations for the soliton parameters. This approach is more simple than the direct perturbation method developed in Refs.~\cite{KonotopVekslerchik_1994, BurtsevCamassa_1997, Lashkin_2004, ChenHuang_1998, AoYan_2005, Ablowitz_2011, Ablowitz_DSModeLockedLasers_2013} for optical dark solitons in several important physical settings connected with linear damping, two-photon absorption, gain with saturation and the effects of the Raman self-induced scattering.

To derive the evolution equations of the soliton parameters, it is possible to use several different but qualitatively similar methods. For example, on substituting~(\ref{eq:AsymptoticSeries}) into~(\ref{eq:psieq}),~(\ref{eq:mReq}), to the first order in $\mu$, we obtain the following linear inhomogeneous differential equation for function $\psi_{1}\!\left(\xi,\mu t\right)$:
\begin{multline}
\label{eq:psi1eq}
\left[iv_{s}\frac{\partial}{\partial\xi}-\frac{1}{2}\frac{\partial^{2}}{\partial\xi^{2}}-1+2\left|\psi_{s}\right|^{2}\right]\psi_{1}+\psi_{s}^{2}\psi_{1}^*=\\ i\frac{\partial v_{s}}{\partial t}\frac{\partial\psi_{s}}{\partial v_{s}}-\mathcal{R}(m^0_{R},\psi_s),
\end{multline}
while perturbation $m^0_{R}\left(\xi,\mu t\right)$ of the polariton density in the reservoir is determined by the wave function $\psi_{s}\!\left(\xi,v_{s}\!\left(\mu t\right)\right)$ of a dark soliton.
It can be proved that spatially localized solutions of Eq.~(\ref{eq:psi1eq}) exist if
\begin{equation}
\label{eq:Fredholm}
\mathrm{Re}\!\left[\,\int\limits_{-\infty}^{+\infty}\!\!\mathrm{d}\xi\Biggl(i\frac{\partial v_{s}}{\partial t}\frac{\partial\psi_{s}}{\partial v_{s}}-\mathcal{R}\left(m^0_R,\psi_{s}\right)\Biggl)\frac{\partial\psi_{s}^{\ast}}{\partial\xi}\right]\!=0,
\end{equation}
which is a full analogue of the Fredholm alternative~\cite{Korn2000, SmirnovMironov_2012}. Ultimately, Eq.~(\ref{eq:Fredholm}) leads to the dynamic equation:
\begin{equation}
\label{eq:Eseq}
\frac{d\mathcal{E}_s}{d t}=v_{s}\int\limits_{-\infty}^{+\infty} d\xi\left(\mathcal{R}\left(m^0_R,\psi_{s}\right)\frac{\partial\psi_{s}^{\ast}}{\partial\xi}+\mathcal{R}^*\left(m^0_R,\psi_{s}\right)\frac{\partial\psi_{s}}{\partial\xi}\right),
\end{equation}
where
\begin{equation}
\label{eq:Es}
\mathcal{E}_{s}=\frac{1}{2}\int\limits^{+\infty}_{-\infty}d\xi\left(\left|\frac{\partial\psi_{s}}{\partial\xi}\right|^{2}+\left(1-\left|\psi_{s}\right|^{2}\right)^{2}\right)=\frac{4}{3}\left(1-v_{s}^{2}\right)^{3/2}.
\end{equation}
is the energy of the dark soliton. Equations (\ref{eq:Eseq}) and (\ref{eq:Es}) fully describe the dynamics (trajectory and velocity) of a 1D dark soliton propagating on a background of a spatially homogeneous polariton condensate.

In essence, Eqs.~(\ref{eq:Eseq}) and~(\ref{eq:Es}) are in agreement with the results of Refs.~\cite{LisakAndersonMalomed_1991, KivsharYang_1994}, and can alternatively be derived using the so-called adiabatic approximation of the perturbation theory for dark solitons.

We stress that our theory so far assumes a perturbative regime of reservoir excitations $m_{R}\!\left(\vec{r},t\right)\!\ll\!1$. We will now show that  this condition is fulfilled in a broad range of regimes.
\subsection{Weak pumping}\label{sec:WeakPump}
First of all, we will consider the regime of a weakly above threshold pump intensity:
\begin{equation}
\label{eq:WeaklyCase}
P_{0}\bigr/P_{th}-1\ll 1.
\end{equation}
The characteristic scale of the perturbation ${m_{R}}_{0}\!\left(\xi,\mu t\right)$ coincides with the dark soliton width $\Lambda_{s}$. Under the condition~(\ref{eq:WeaklyCase}), $\bar{\gamma}_{R}\!\gg\!1$, and, thereby, for all velocities $v_{s}$, $\bar{\gamma}_{R} \gg 1/\Lambda_{s}$. According to Eq.~(\ref{eq:mReq}) and the asymptotic series~(\ref{eq:AsymptoticSeries}), the perturbation $m^0_R \left(\xi,\mu t\right)$ of the polariton density in the reservoir is coupled to $\left|\psi_{s}\!\left(\xi,v_{s}\!\left(\mu t\right)\right)\right|^{2}$:
\begin{equation}
\label{eq:mR0:WeakPump}
m^0_{R}=\frac{\bar{\gamma}_{C}}{\bar{\gamma}_{R}}\Bigl(1-\left|\psi_{s}\right|^{2}\Bigr),
\end{equation}
and is a small value, provided that $\bigl.\bar{\gamma}_{C}\bigr/\bar{\gamma}_{R} \equiv \gamma_{C}\bigr/\gamma_{R}\!\ll\!1$. Hence, the role of the small parameter in our problem is now played by the ratio $\mu \sim \gamma_{C}/\gamma_{R} \ll\!1$.
Substituting~(\ref{eq:mR0:WeakPump}) into the right-hand-side of equality~(\ref{eq:Eseq}), taking into account that $\psi_{s}\!\left(\xi,v_{s}\right)$ is determined by~(\ref{eq:ds}), and integrating, we obtain the expression for the dark soliton acceleration
\begin{equation}
\label{eq:vseq:WeakPump}
\frac{d v_{s}}{ d t}=-\frac{v_{s}\left(1-v_{s}^{2}\right)}{2\tau_{1}},
\end{equation}
and velocity:
\begin{equation}
\label{eq:vs:WeakPump}
v_{s}^{2}\!\left(t\right)=\frac{{v_{s}^{2}}_{0}\exp\!\left(t\bigl/\tau_{1}\bigr.\right)}{1-{v_{s}^{2}}_{0}+{v_{s}^{2}}_{0}\exp\left(t\bigl/\tau_{1}\right)}.
\end{equation}
The velocity of the dark soliton $v_{s}$ determines its ``darkness'' (contrast) through the simple relation ${n^{\rm min}_{C}}\left(t\right)=v_{s}^{2}\!\left(t\right)$. The Eq.~(\ref{eq:vs:WeakPump}) shows that $n^{\rm min}_{C}$ grows in time approaching the unit background, and as a result  the dark soliton disappears at the time:
\begin{equation}
\label{eq:tau_WeakPump}
\tau_{1}=\frac{3}{2}\frac{\bar{\gamma}_{R}}{\bar{R}\bar{\gamma}_{C}}=\frac{3}{2}\frac{1}{\gamma_{C} \tau_0}\frac{P_{th}}{P_0-P_{th}},
\end{equation}
where $\tau_0$ is the characteristic time scale for our system introduced in Sec. \ref{sec:Model}. We point out that this case is similar to the NLS model for a defocusing nonlinear medium with saturated gain analyzed in Ref.~\cite{ KivsharYang_1994}.

\begin{figure}[htb]
\includegraphics[width=8.5cm]{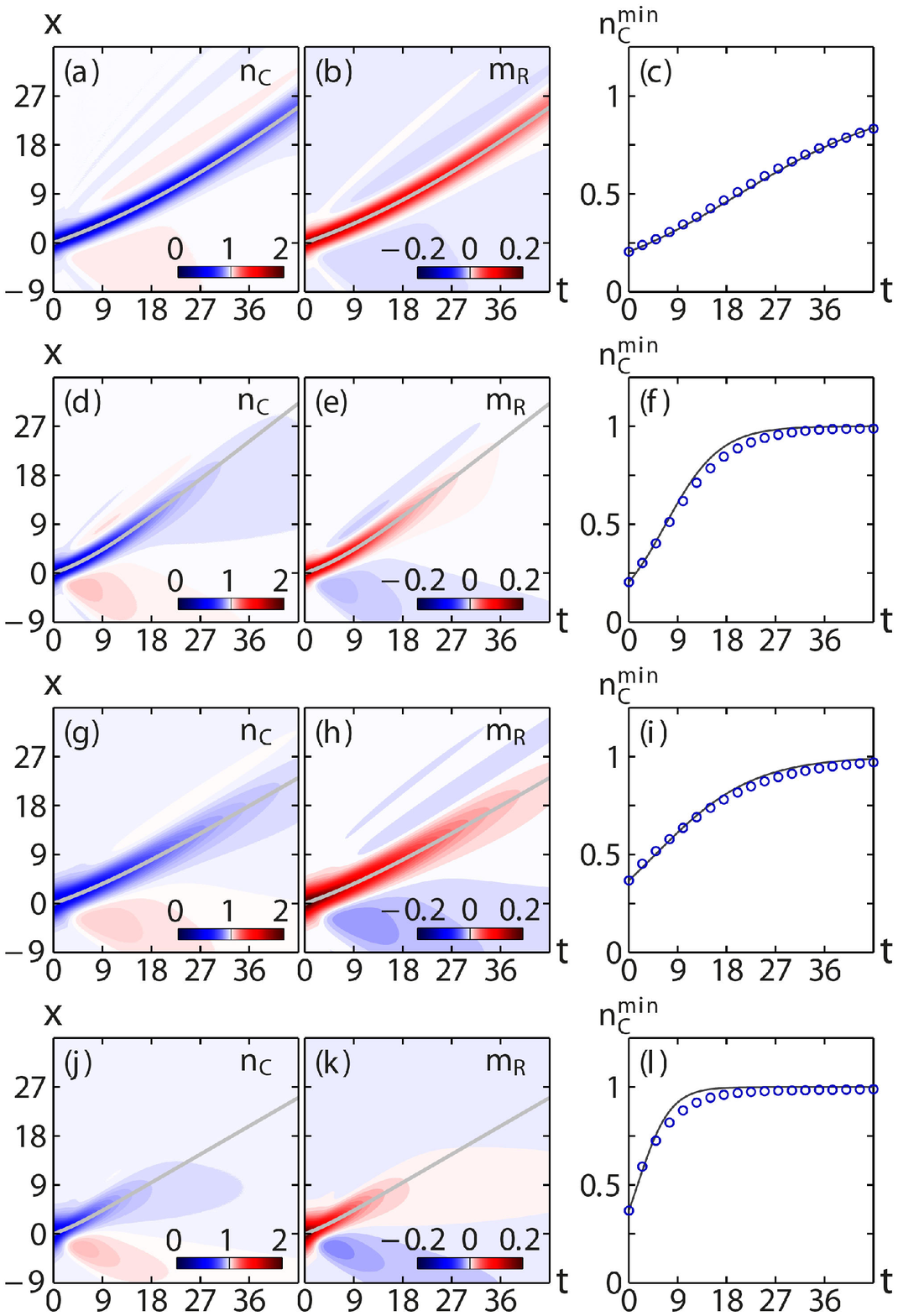}
\caption{Dynamics of a 1D dark soliton with the initial velocity $v_s(0)=0.35$ in the case of weak pumping. Shown are contour plots of $n_{C}\left(x,t\right)$ (left column) and $m_{R} \left(x,t\right)$ (middle column), and dependence $n^{\rm min}_C(t)$ (right column, circles) computed using Eqs.~(\ref{eq:psieq}) and (\ref{eq:mReq}). Solid lines are calculated using analytical formulas Eq.~(\ref{eq:vseq:WeakPump}) (left and middle columns) and Eq.~(\ref{eq:vs:WeakPump}) (right column).  Parameters are $\bar{g}_R=2$, $\bar{\gamma}_C=3$, and (a-c) $\bar{\gamma}_{R}=15$, $\bar{R}=0.5$; (d-f) $\bar{\gamma}_{R}=15$, $\bar{R}=1.5$;  (g-i) $\bar{\gamma}_{R}=9$, $\bar{R}=0.5$; (j-l) $\bar{\gamma}_{R}=9$, $\bar{R}=1.5$.} \label{fig:WeakPump_1}
\end{figure}

\begin{figure}[h]
\includegraphics[width=8.5cm]{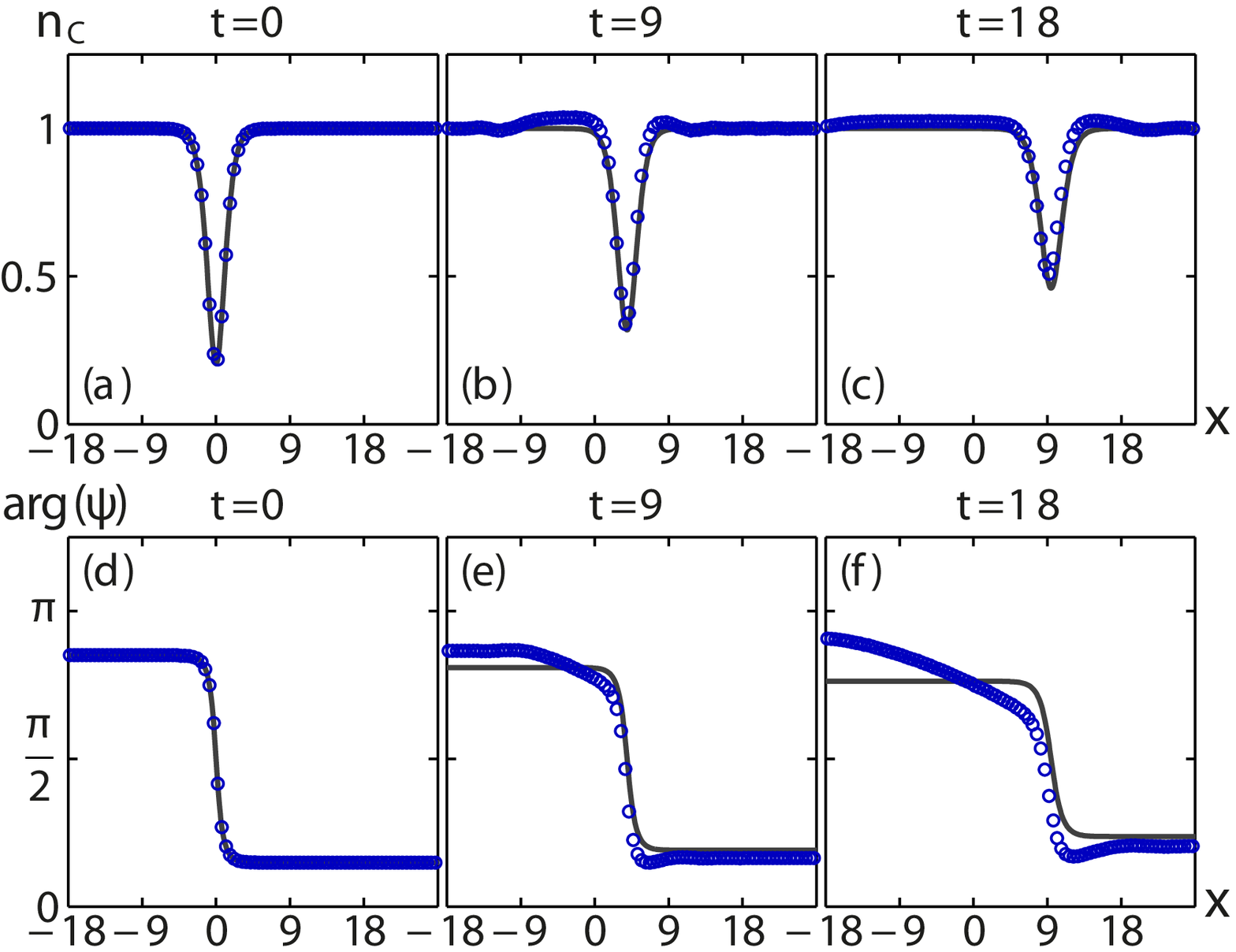}
\caption{\label{fig:relaxation:weakP} Cross-section of a 1D dark soliton (a-c) density and (d-f) phase for different stages of relaxation dynamics shown in Fig. \ref{fig:WeakPump_1}(a-c). Solid lines are obtained analytically. }
\end{figure}


As seen form Eqs.~(\ref{eq:vseq:WeakPump}),~(\ref{eq:tau_WeakPump}) and~(\ref{eq:vs:WeakPump}), soliton relaxation process is not affected by the interaction with reservoir. It can be easily shown by substituting expression~(\ref{eq:mR0:WeakPump}) for $m_R=m^0_R$ into Eq.~(\ref{eq:psieq}). Hereby, Eq.~(\ref{eq:psieq}) takes the form:
\begin{equation}
\label{eq:modifiedNLS}
i\frac{\partial\psi}{\partial t}+\frac{1}{2}\Delta\psi+\alpha\Bigl(1-\left|\psi\right|^{2}\Bigr)\psi=\frac{i}{2}\bar{R}m_{R}\psi
\end{equation}
where $\alpha=1-\bar{g}_{R}\bar{\gamma}_{C}\bigl/\bar{\gamma}_{R}$. According to~(\ref{eq:modifiedNLS}), interaction with the reservoir, characterized by $\bar{g}_{R}$, only slightly changes the width and darkness of the solitonic state, not affecting its acceleration, which is determined by the stimulated scattering term proportional to $\bar{R}$. We point out that at $\bar{g}_{R}\bar{\gamma}_{C}\bigl/\bar{\gamma}_{R}>1$ the sign of $\alpha$ is changed to the opposite, i.e. the character of nonlinearity is effectively switched from defocusing to {\em focusing}. In this regime, the condensate can become unstable.

To compare our analytical  predictions with direct numerical simulation of the model equations, we show the evolution of the condensate density distribution $n_{C}\!\left(x,t\right)\!=\!\left|\psi\!\left(x,t\right)\right|^{2}$ and associated perturbations of the polariton reservoir density $m_{R}\!\left(x,t\right)$ in Fig. \ref{fig:WeakPump_1} (left and middle columns, respectively). For comparison, the dark soliton trajectory calculated using Eq.~(\ref{eq:vseq:WeakPump}) is depicted by the solid line and demonstrates a remarkable agreement with the results of direct integration of Eqs.~(\ref{eq:psieq}),~(\ref{eq:mReq}). The right column in Fig. \ref{fig:WeakPump_1} shows the time-dependence of the minimum value of the condensate density associated with the dark soliton. The solid line shows the darkness ${n^{\rm min}_{C}}\left(t\right)=v_{s}^{2}\!\left(t\right)$ calculated analytically using Eq.~(\ref{eq:vs:WeakPump}), and shows an excellent agreement with numerics. As seen in Fig. \ref{fig:WeakPump_1} (d-f), the soliton lifetime,  $\tau_{1}$, reduces dramatically with growth of the stimulated scattering rate $\bar{R}$, characterizing the efficiency of the polariton scattering into the condensed state. Namely, the three-fold increase in $\bar{R}$, compared to the parameters in Figs.~\ref{fig:WeakPump_1} (a-c), leads to the three-fold decrease in $\tau_{1}$, in agreement with Eq.~(\ref{eq:tau_WeakPump}). Likewise, longer lifetime of the reservoir polaritons leads to shorter lifetime of dark solitons. As can be seen in Figs.~\ref{fig:WeakPump_1}(g-i), the relaxation time $\tau_{1}$ has decreased by $5/3$ when $\bar{\gamma_R}$ decreased by $2/5$, in full compliance with~(\ref{eq:tau_WeakPump}). At this ratio $\bar{\gamma}_{C}\bigl/\bar{\gamma}_{R}=1/3$, therefore for correct description of the dark soliton dynamics in the zero order the wave function $\psi\!\left(x,t\right)$ should be set taking into account the coefficient $\alpha$ [see Eq.~(\ref{eq:modifiedNLS})], which determines the soliton width $\Lambda_{s}=\left(\alpha-v_{s}^{2}\right)^{-1/2}$ and the density minimum $n^{\rm min}_C=v_{s}^{2}/\alpha$. Conversely, simultaneous increase in both radiative and non-radiative decay rates of the reservoir, $\bar{\gamma}_{R}$ and $\bar{R}$, causes the solitonic structures to disappear rather fast. In the particular case shown in Figs.~\ref{fig:WeakPump_1} (j-l), the dimensionless relaxation time is very short $\tau_{1}=3$, as predicted by Eq.~(\ref{eq:tau_WeakPump}). However, even in this case, our asymptotic description gives quite satisfactory understanding of the behavior of the dark localized structures at all stages of evolution. Although neither the initial density nor the phase structure typical of a dark soliton in a conservative system survives in the open-dissipative condensate, the localized nature of the dark solitonic excitation with the distinct phase gradient across its profile is preserved during the relaxation process (see Fig. \ref{fig:relaxation:weakP}).

Finally, we note that, if we assume the polariton relaxation time $\gamma^{-1}_c=10$ ${\rm ps}$ and the dimensionless parameters in Figs.~\ref{fig:WeakPump_1}(a-l), the time scaling variable expressed as $\tau_0=\bar{\gamma_c}/\gamma_c$ takes the physical value of $30$ ${\rm ps}$. The corresponding propagation time for the solitonic state shown in Figs. ~\ref{fig:WeakPump_1} (a-c) reaches $t=1800$ ${\rm ps}$, which is much longer than condensate and reservoir relaxation times.

\subsection{Slow solitons}\label{sec:SmallVelocities}
If the velocity of a dark soliton is smaller than all characteristic relaxation and scattering rates in the system, i.e.,
\begin{equation}
\label{eq:SmallVelocities}
v_{s}\ll\bar{\gamma}_{C},\bar{\gamma}_{R},\bar{R},
\end{equation}
we can repeat the analysis described above. Under the condition~(\ref{eq:SmallVelocities}), the perturbation $m^0_R\left(\xi,\mu t\right)$ of the reservoir density depends on $\left|\psi_{s}\left(\xi,v_{s}\left(\mu t\right)\right)\right|^{2}$ as:
\begin{equation}
\label{eq:mR0:SmallVelocities}
m^0_R=\frac{\bar{\gamma}_{C}\left(1-\left|\psi_{s}\right|^{2}\right)}{\bar{\gamma}_{R}+\bar{R}\left|\psi_{s}\right|^{2}},
\end{equation}
and is a small value, provided that $\bar{\gamma}_{C}/\left(\bar{\gamma}_{R}+\bar{R}\right) \ll 1$. The relation $\bar{\gamma}_{C}/\left(\bar{\gamma}_{R}+\bar{R}\right)$ now plays the role of a small parameter of the problem $\mu \sim \bar{\gamma}_{C}/\left(\bar{\gamma}_{R}+\bar{R}\right)\ll1$.
Substituting~(\ref{eq:mR0:SmallVelocities}) into the right-hand-side of Eq.~(\ref{eq:Eseq}), taking into account the expression for $\psi_{s}\left(\xi,v_{s}\right)$  (\ref{eq:ds}), and integrating, we obtain the following equation for the acceleration:
\begin{equation}
\label{eq:vseq:SmallVelocities}
\frac{d v_s}{d t}=\frac{1}{\tau_2}v_s,
\end{equation}
and the soliton velocity:
\begin{equation}
\label{eq:vs:SmallVelocities}
v_{s}^{2}\!\left(t\right)={v_{s}^{2}}_{0}e^{t\bigl/\tau_{2}}.
\end{equation}
Here the relaxation time is:
\begin{multline}
\label{eq:tau_SmallVelocities}
\tau_{2}=\frac{\bar{\gamma}_{C}}{2}\left[\frac{\bar{\gamma}_{R}+\bar{R}}{\sqrt{\bar{R}\bar{\gamma}_{R}}}\arctan \sqrt{\frac{\bar{R}}{\bar{\gamma}_{R}}}-1\right]\approx\\ 
\frac{1}{2} \gamma_c \tau_0 \frac{P_{th}}{P_0-P_{th}}. 
\end{multline}
From Eq. (\ref{eq:vs:SmallVelocities}) it follows that a stationary {\em black} soliton with $v_{s}(0)=0$ is {\em unstable}. Any small perturbation of the initial velocity leads to soliton acceleration, which results in the exponential growth of the velocity. When the condition $v_s\ll1$ is violated, acceleration slows down and soliton behaves as described in the previous Section, losing its energy and relaxing to the background. Rapid relaxation of quasi-stationary dark solitons was numerically demonstrated in Ref. \cite{DSinc2013} for moderate pumping intensities.
\subsection{Strong pumping}\label{sec:StrongPumping}
The above results were obtained in the weak pump approximation. However, we can also use our general asymptotic approach to analyze the case when the pump is strong:
\begin{equation}
\label{eq:StrongPump}
P_{0}/P_{th}\gg 1.
\end{equation}
Taking into account the asymptotic expansion~(\ref{eq:AsymptoticSeries}) to the first order in $\mu$, Eq.~(\ref{eq:mReq}) can be rewritten in the form:
\begin{equation}
\label{eq:mR0eq}
-v_{s} \frac{\partial m^0_R}{\partial\xi}+\left(\bar{\gamma}_{R}+\bar{R}\left|\psi_{s}\right|^{2}\right)m^0_R=\bar{\gamma}_{C}\left(1-\left|\psi_{s}\right|^{2}\right).
\end{equation}

\begin{figure}[hb]
\includegraphics[width=8.5cm]{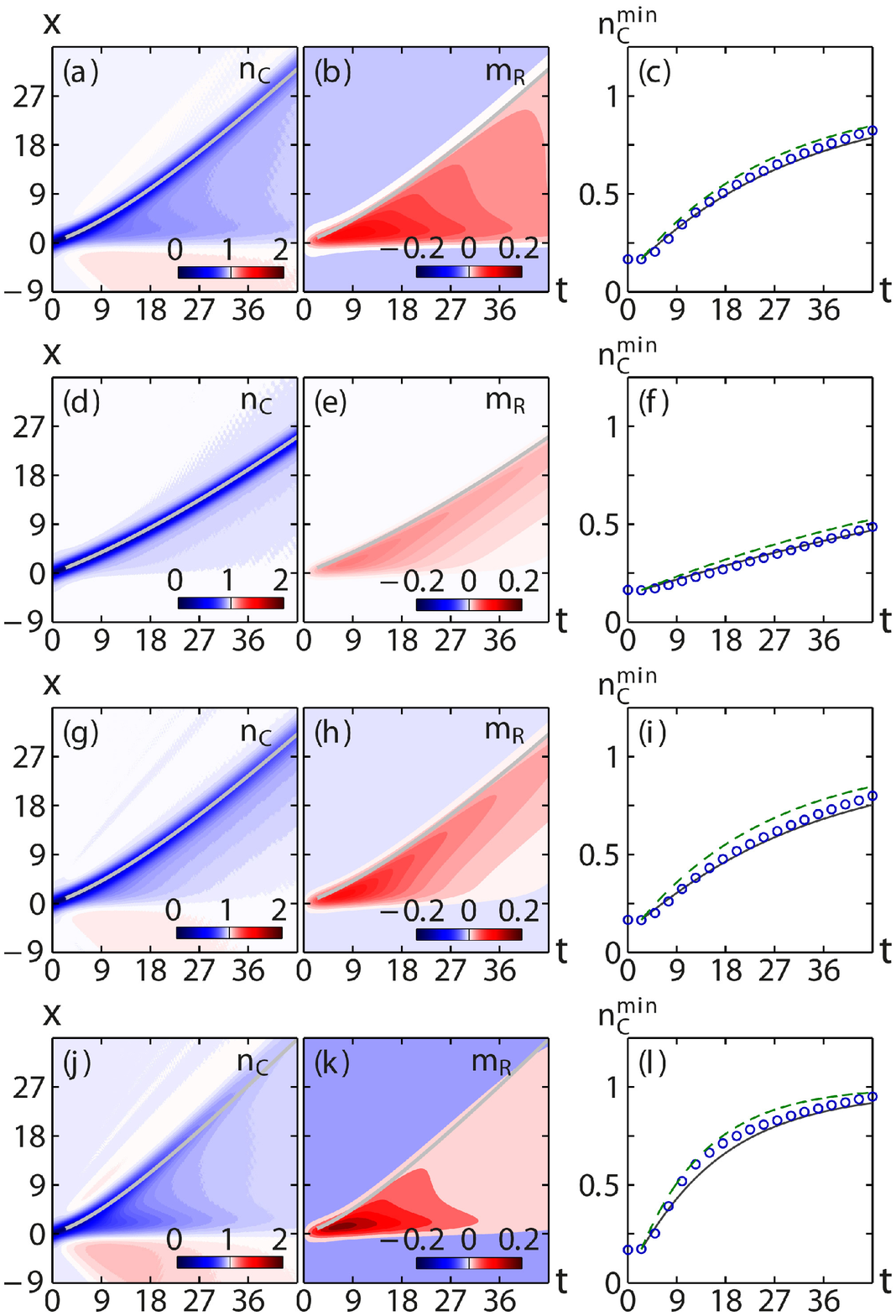}
\caption{\label{fig:StrongPump_1} Dynamics of a 1D dark soliton with the initial velocity ${v_{s}}(0)=0.4$ in the case of strong pumping. Shown are contour plots of $n_{C}\!\left(x,t\right)$ (left column) and $m_{R}\!\left(x,t\right)$ (middle column), and the dependence  $n^{\rm min}_{C}(t)$ (right column, circles) obtained by numerical solution of Eqs.~(\ref{eq:psieq}) and (\ref{eq:mReq}). The solid lines in the left and middle columns are obtained using Eq.~(\ref{eq:vseq:StrongPump_a}). Solid line in the right column is obtained analytically using the approximate Eq.~(\ref{eq:vs:StrongPump}), for comparison the dashed line is calculated by numerical integration of Eq. (\ref{eq:vseq:StrongPump_a}). Parameters are: (a-c) $\bar{\textsl{g}}_{R}=2$, $\bar{\gamma}_{C}=0.03$, $\bar{\gamma}_{R}=0.05$, $\bar{R}\!=\!0.04$; (d-f) $\bar{\gamma}_{C}\!=\!0.01$, $\bar{\gamma}_{R}\!=\!0.05$, $\bar{R}\!=\!0.04$; (g-i) $\bar{\gamma}_{C}\!=\!0.03$, $\bar{\gamma}_{R}\!=\!0.07$, $\bar{R}\!=\!0.08$; (j-l) $\bar{\gamma}_{C}\!=\!0.06$, $\bar{\gamma}_{R}\!=\!0.04$, $\bar{R}\!=\!0.15$. }
\end{figure}

Therefore, the perturbation of the polariton density in the reservoir depends on the wave function nonlocally:
\begin{equation}
\label{eq:mR0:StrongPump_a}
m^0_{R}=\wp e^{\left(p\eta-q\tanh \eta \right)}\int\limits_{\eta}^{+\infty}e^{\left(q\tanh \eta'-p \eta'\right)}{\rm sech}^{2}\eta' {\rm d}\eta',
\end{equation}
where 
\begin{eqnarray}
\eta=\sqrt{1-v_{s}^{2}}\xi,\,\,
\wp=\frac{\sqrt{1-v_{s}^{2}}}{v_{s}}\bar{\gamma}_{C},\\  \nonumber
p=\frac{\left(\bar{\gamma}_{R}+\bar{R}\right)}{v_{s}\sqrt{1-v_{s}^{2}}},\,\,
q=\frac{\sqrt{1\!-\!v_{s}^{2}}}{v_{s}}\bar{R}. \nonumber
\end{eqnarray}
Under the condition~(\ref{eq:StrongPump}), $\bar{\gamma}_{R}\!\ll\!1$ and $\bar{\gamma}_{C}\ll1$ and thereby the value of $m^0_R \left(\eta\right)$ determined by (\ref{eq:mR0:StrongPump_a}) is small in a wide range of values of the soliton velocity $v_{s}$. Here, $\wp$ is a small parameter of the system $\mu\sim\wp \ll 1$.

If $p \lesssim 1$, that is $\bar{R} \lesssim v_{s}\sqrt{1 - v_{s}^{2}}$, we can replace $p \eta'$ by $p\tanh \left(\eta' \right)$ in~(\ref{eq:mR0:StrongPump_a}) and obtain an analytical expression for the function $m^0_R\left(\eta\right)$:
\begin{equation}
\label{eq:mR0:StrongPump_b}
m^0_R=\frac{\wp}{\left(p-q\right)}e^{\left(p\eta-q\tanh \eta \right)}\left[e^{\left(p-q\right)}-e^{[\left(p-q\right)\tanh \eta ]}\right].
\end{equation}\par
Substituting~(\ref{eq:mR0:StrongPump_b}) into the right-hand-side of Eq.~(\ref{eq:Eseq}) and taking into account the expresson for  $\psi_{s}\!\left(\xi,v_{s}\right)$~(\ref{eq:ds}), we derive the following expression for the soliton acceleration:
\begin{multline}
\label{eq:vseq:StrongPump_a}
\frac{d v_s}{d t}=-\frac{\bar{\gamma}_C}{4}\frac{\sqrt{1-v^2_s}}{w^2}\Bigl[\frac{\bar{g}_{R}\left(\bar{\gamma}_{R}+\bar{R}\right)}{v_{s}^{2}}\left(e^{-2w}+\frac{2}{3}w-1\right)+ \\
2\bar{g}_R\bar{R}\frac{\left(1-v_{s}^{2}\right)}{v_s^2w^2}\left(\left(w+1\right)e^{-2w}+w-1\right)- \\
\bar{R}\left(e^{-2w}+2w-1\right)\Bigr],
\end{multline}
where
\begin{equation}
w=\frac{\bar{\gamma}_{R}+\bar{R}v_{s}^{2}}{v_{s}\sqrt{1-v_{s}^2}}.
\end{equation}
Note that $w\!\left(v_{s}\right)$ can be small if $\bar{R}v_{s}\ll\sqrt{1\!-\!v_{s}^{2}}$, and this is possible even for $\bar{R}\!\sim\!v_{s}$. When $w\!\left(v_{s}\right)\!\ll\!1$, Eq.~(\ref{eq:vseq:StrongPump_a}) can be reduced to:
\begin{equation}
\label{eq:vseq:StrongPump_b}
\frac{dv_{s}}{d t}=-\frac{d v_s}{d t}=\frac{1}{3\tau_{3}}\frac{\left(1-v_{s}^{2}\right)}{v_{s}},
\end{equation}
where
\begin{equation}
\label{eq:tau_StrongPump}
\tau_{3}=\frac{3}{2}\frac{1}{\bar{\gamma}_{C}\bar{\textsl{g}}_{R}}=\frac{3}{2}\frac{1}{\gamma_c\gamma_R\tau^2_0}\frac{P_{th}}{\left(P_{0}-P_{th}\right)}
\end{equation}
is the the soliton relaxation time. By integrating~(\ref{eq:vseq:StrongPump_b}), we obtain the expression for the soliton velocity:
\begin{equation}
\label{eq:vs:StrongPump}
v_{s}^{2}\!\left(t\right)=1-\left(1-{v_{s}^{2}}_{0}\right) e^{-t\bigl/\tau_{3}}.
\end{equation}
As seen from~(\ref{eq:vseq:StrongPump_b}),~(\ref{eq:tau_StrongPump}), the dominant contribution to the perturbation ~(\ref{eq:RightHandSide}) causing relaxation of a dark soliton is $\bar{g}_{R}m_{R}\psi$. This reflects the nonlocal coupling between $\left|\psi_{s}\right|^{2}$ and $m^0_{R}$. In the case of strong pumping (in contrast to that of a weak pump) the lifetime of the solitonic structure is determined by the nonlinear interaction of condensate and reservoir polaritons. In all our calculations we took $\bar{g}_{R}=g_{R}/g_{C}=2$, as predicted by the Hartree-Fock theory. However, the strength of the condensate-reservoir interaction $\bar{\textsl{g}}_{R}$ has not been conclusively verified in experiments. Therefore, dynamics of dark solitons created in an experiment via strong incoherent excitation, would not only serve as a test for our model, but may potentially lead to accurate estimation of $\bar{g}_R$ from the characteristic time of relaxation $\tau_{3}$.

\begin{figure}[t]
\includegraphics[width=8.5cm]{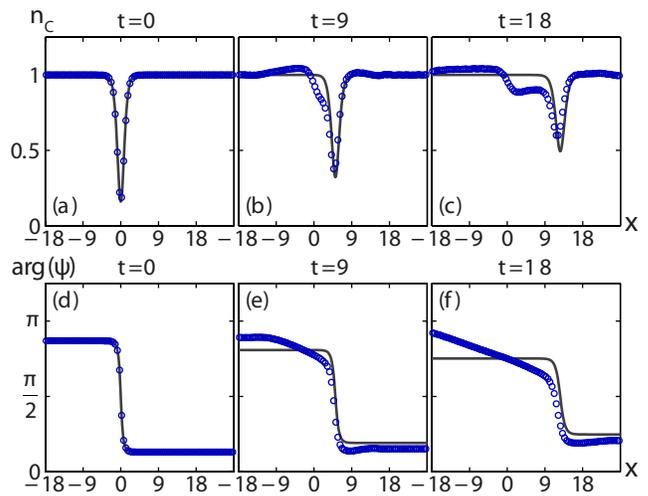}
\caption{\label{fig:relaxation:strongP} Cross-section of a 1D dark soliton (a-c) density and (d-f) phase for different stages of relaxation dynamics shown in Fig. \ref{fig:StrongPump_1}(a-c). Solid lines are obtained analytically. }
\end{figure}

To demonstrate agreement between our analytical theory and numerical simulations, we show evolution of the polariton density distribution $n_{C}\!\left(x,t\right)\!=\!\left|\psi\!\left(x,t\right)\right|^{2}$ and associated perturbations of the reservoir density $m_{R}\!\left(x,t\right)$, in the left and middle column panels of Fig.\ref{fig:StrongPump_1}, respectively, for various values of the parameters. In contrast to  Fig.~\ref{fig:relaxation:weakP}, nonlocal coupling between the reservoir density $m_{R}\left(x,t\right)$ and the condensate density $n_{C}\left(x,t\right)$ significantly affects the soliton dynamics, as seen in Fig.~\ref{fig:relaxation:strongP}. The reservoir exerts a "drag" on the dark soliton, whereby the soliton delocalizes and develops a low density tail [Fig.~\ref{fig:relaxation:strongP}(c)].  Nevertheless, the density minimum propagates in remarkable agreement with the the dark soliton trajectory calculated using Eq.~(\ref{eq:vseq:StrongPump_a}) as depicted by the solid line in Figs.~\ref{fig:relaxation:strongP}(a,d,h,j) and (b,e,h,k). Time-dependence of the minimum density value, ${n^{\rm min}_{C}}\left(t\right)=v_{s}^{2}\left(t\right)$, calculated using Eq.~(\ref{eq:vs:StrongPump}) is also in excellent agreement with the numerical solutions of Eqs.~(\ref{eq:psieq}) and (\ref{eq:mReq}), as shown in Figs.~\ref{fig:StrongPump_1} (c,f,i,l). 

In physical terms, stronger pumping leading to larger densities of the condensate {\em cw} background, shortens the time scale of the soliton relaxation dynamics. In particular, the dynamics shown in Fig. \ref{fig:StrongPump_1} is very fast, and its time scale is comparable to the polariton decay times. For $\gamma^{-1}_c=10$ ${\rm ps}$, corresponding propagation times are: $t=18$ ${\rm ps}$ (a-c,g-i), $t=6$ $ps$ (d-f), and $t=36$ ${\rm ps}$ (j-l).

\section{Transverse instability of dark solitons}\label{sec:DSInstability}
It is well known that, in two (or three) dimensions, quai-one-dimensional dark soliton stripes described by solutions~(\ref{eq:ds}) of NLS equation~(\ref{eq:NLS}) are unstable to perturbations of sufficiently long wavelength in direction transverse to the soliton line (or plane) ~\cite{KuznetsovTuritsyn_1988}. The {\em transverse instability} of dark solitons has been extensively discussed in the literature on the NLS model ~(\ref{eq:NLS}) in the context of media with defocusing nonlinearity, such as optical media with a nonlinear negative correction to the refractive index~\cite{Opt.Lett._1993, Opt.Commun._1993, Europhys.Lett._1995, PhysRevE.51.5016_1995, Phys.Rep._2000} and atomic condensates~\cite{Analogies_2004, KamchatnovPitaevskii_ConvectiveInstability_2008, Kamchatnov_ConvectiveInstability_2011, HoeferIlan_2012, MironovSmirnov_2011}. Experiments\cite{Tikhonenko_1996, PhysRevLett.76.2262_1996, PhysRevA.54.870_1996, PhysRevLett.86.2926_2001} confirm that quasi-one-dimensional nonlinear structures in these media are unstable to transverse modulations and may decay into pairs of vortices with opposite topological charges. 

In the context of exciton-polariton physics, dark soliton stripes emerged as an important topic in connection with the physics of superfluidity and breakdown of the superfluid flow in a coherently driven condensate \cite{DSInstability_PolaritonBEC_2011, DSInstability_PolaritonBEC_2012}. Here, based on the ideas detailed in Sec.~\ref{sec:DSDynamic}, we consider peculiarities of transverse instability development of dark solitons in a condensate of polaritons with an {\em incoherent} pump. This is a regime that so far has not been studied in experiments, and therefore our predictions could be used to guide further experimental efforts.

Dark soliton stripes in conservative condensates and optical fields are always unstable to transverse long-wavelength modulations, and the linear stage of this instability was studied in Refs.~\cite{KuznetsovTuritsyn_1988, KamchatnovPitaevskii_ConvectiveInstability_2008, Kamchatnov_ConvectiveInstability_2011, HoeferIlan_2012}. The linear stability of dark solitons may be analyzed by means of the Bogoliubov-de Gennes method~\cite{PitaevskiiStringari2003, PethickSmith2008} by introducing small perturbations about the dark soliton~(\ref{eq:ds}) in the manner described in Sec. \ref{BdG}. For a dark soliton in a conservative system described by Eq.~(\ref{eq:NLS}), the structure of the excitations spectrum $\omega \left(k\right)$ can be computed numerically for any given values of the soliton velocity~$v_{s}$, and is well known. The spectrum contains both continuous and discrete parts, with exactly two purely imaginary discrete eigenvalues $\pm i\Gamma\!\left(k\right)$ emerging for  $0\!<\!\left|k\right|\!<\!k_{c}$, where
\begin{equation}
k=k_{c}\!\left(v_{s}\right)=\left[2\sqrt{v_{s}^{4}-v_{s}^{2}+1}-v_{s}^{2}-1\right]^{1/2}.
\end{equation}
Therefore, the dark soliton is unstable to transverse perturbations of sufficiently long wavelength. Beyond demonstrating the existence of an instability, knowledge of the dispersion relation $\omega\!\left(k\right)$ for a range of wavenumbers $k$ yields important properties of the instability, such as the maximum growth rate $\Gamma_{max}$ and the associated wavenumber $k_{max}$ of maximally unstable transverse perturbations.
 
There is no analytical expression for the full instability spectrum, although its approximations ~\cite{PhysRevE.51.5016_1995, KamchatnovPitaevskii_ConvectiveInstability_2008, HoeferIlan_2012} and asymptotic expressions  \cite{KuznetsovTuritsyn_1988, Phys.Rep._2000, KamchatnovPitaevskii_ConvectiveInstability_2008} are known. Evolution of shallow (low contrast) dark solitons ($1-v_{s}^{2} \ll 1$) is asymptotically described by the Kadomtsev-Petviashvilli equation~\cite{KuznetsovTuritsyn_1988, Phys.Rep._2000, SmirnovMironov_2012}, and the dispersion law in this limit can be obtained in the explicit form~\cite{Zakharov_1975}. For shallow solitons it is also possible to determine the maximum growth rate $\Gamma_{max}$ and maximally unstable wavenumber $k_{max}$ asymptotically~\cite{HoeferIlan_2012}.

\begin{figure}[htb]
\includegraphics[width=8.5cm]{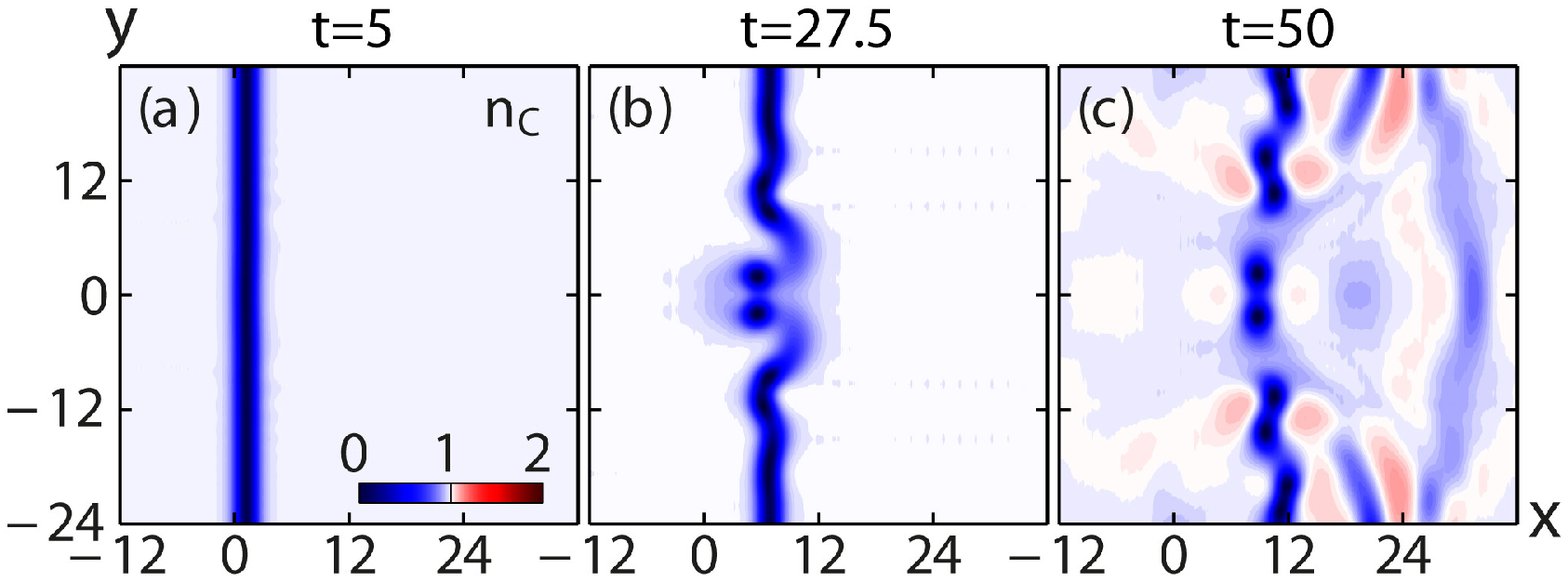}
\caption{\label{fig:NLS_1} Snapshots at (a) $t=5$, (b) $t\!=\!27.5$, (c) $t\!=\!50$ of transverse instability development of a dark soliton solution of NLS equation~(\ref{eq:NLS}). The initial velocity $v_{s}(t=0)$ is modulated along $y$-axis by the Gaussian function \ref{eq:NumInitDistribution}. Parameters are ${v_{s}}_{0}\!=\!0.25$, $\delta {v_{s}}_{0}\!=\!0.025$ and $\sigma_{y}\!=\!1$. Corresponding maximum increment of transverse instability is $\Gamma_{max} \approx 0.2293$.}
\end{figure}

%

The nonlinear stage of the transverse instability development within the framework of Eq.~(\ref{eq:NLS}) was studied both numerically~\cite{Opt.Lett._1993, Opt.Commun._1993, Europhys.Lett._1995, PhysRevE.51.5016_1995, KamchatnovPitaevskii_ConvectiveInstability_2008, Kamchatnov_ConvectiveInstability_2011, MironovSmirnov_2011, HoeferIlan_2012} and analytically~\cite{PhysRevE.51.5016_1995, Phys.Rep._2000, MironovSmirnov_2011}. A density drop in a conservative condensate does not simply disappear as a result of dark soliton instability~\cite{KuznetsovTuritsyn_1988}. As was shown in Ref.~\cite{MironovSmirnov_2011} a solitonic structure, which locally resembles~(\ref{eq:ds}), is spatially localized at each moment $t$ near a smooth reference curve. In the process of the soliton decay, the contrast of the density drop reaches its maximum value (density drops to zero) with a collapse in some regions of this curve. Topological defects, i.e., vortices, are generated at the points on the reference line where both the condensate density and the curvature of the line reach zero. Ultimately, a family of moving vortex-antivortex pairs is formed \cite{MironovSmirnov_2011}. 
If the initial velocity of a moving dark (gray) soliton approaches the velocity of sound in the condensate, the nonlinear instability stage terminates in the formation of {\em vortex-free} structures resembling two-dimensional {\em Kadomtsev-Petviashvilli solitons}~\cite{PhysRevE.51.5016_1995, Phys.Rep._2000, MironovSmirnov_2011}. Figure~\ref{fig:NLS_1} demonstrates the typical development of transverse modulational instability of the dark soliton with the velocity (width) slightly modulated at $t=0$ by the Gaussian function:
\begin{eqnarray}
\label{eq:NumInitDistribution}
\psi\!\left(\vec{r},t\!=\!0\right)=\psi_{s}\!\left(x,v_{s}\!\left(y\right)\right),\\ \nonumber
v_{s}\!\left(y\right)={v_{s}}_{0}+\delta {v_{s}}_{0}e^{-y^2\bigl/\sigma_{y}^{2}}.
\end{eqnarray}
The calculations were performed in the framework of Eq.~(\ref{eq:NLS}). As seen from the snapshots of the density distribution $\left|\psi\right|^{2}$, in the domain of modulation instability the quasi-1D solitonic structure ~(\ref{eq:NumInitDistribution}) is destroyed and gives rise to the first vortex-antivortex pair at $t=27.5$. At the vortex core, the density $\left|\psi\right|^{2}$ of the condensate is zero. Then, at $t=50$, two more vortex pairs appear, and the process repeats at the later stages of the instability development (hence the term  "snake" instability). As a result, the soliton breaks up into several interacting vortex-antivortex pairs. Increasing the initial velocity slows down the process of vortex pairs formation.

\begin{figure}[t]
\includegraphics[width=8.5cm]{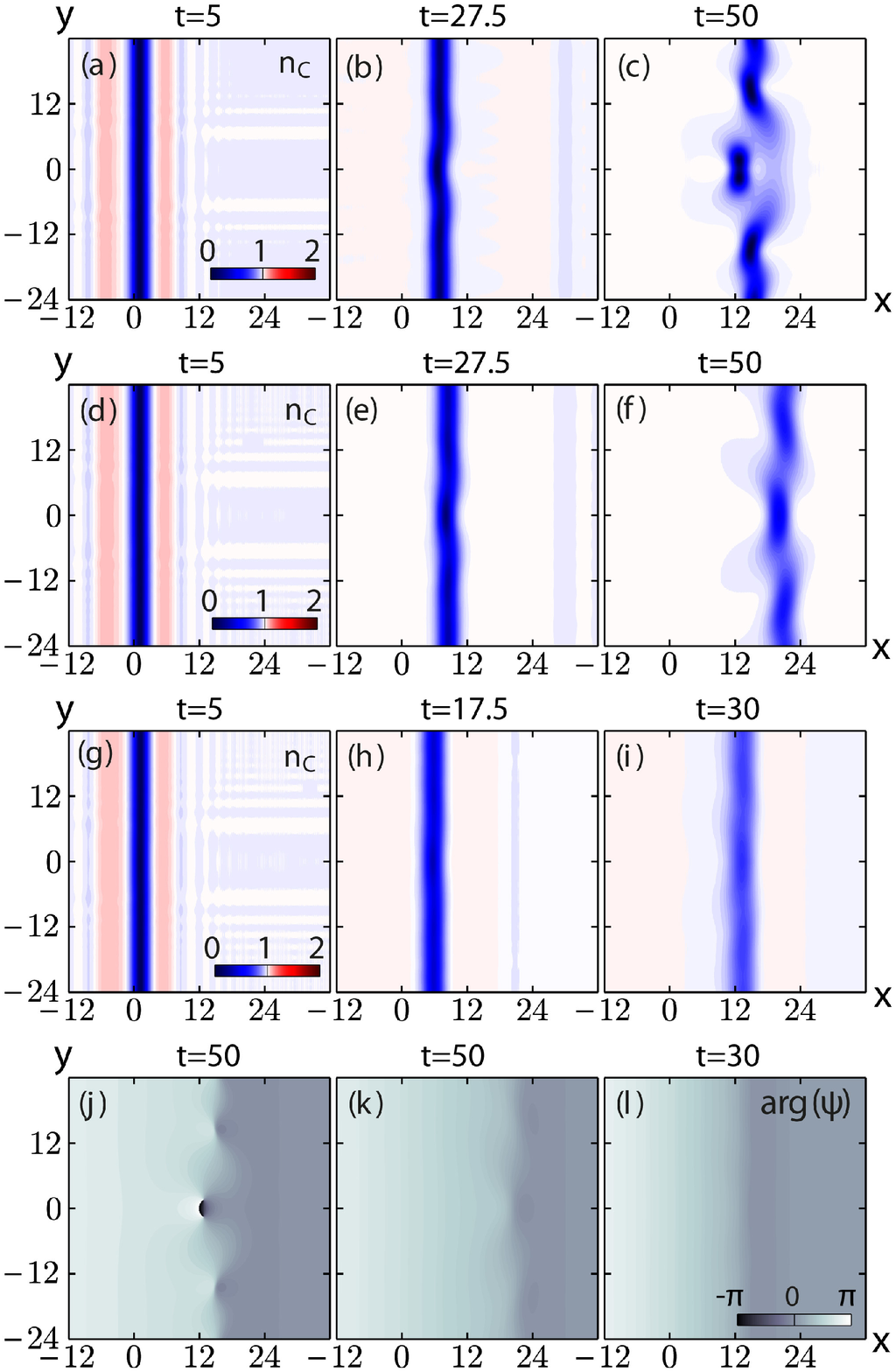}
\caption{\label{fig:TrInstability_WeakPump_1} Transverse instability development of a dark soliton with the initial velocity ${v_{s}}_{0}=0.25$ for weak pumping.  The initial condensate wave function $\psi\!\left(\vec{r},t\right)$ is the same as that in Fig.~\ref{fig:NLS_1} and $m_{R}\!\left(\vec{r},0\right)=0$. Shown are: (a-i) condensate density $n_C$, and (j-l) phase ${\rm arg}(\psi)$ corresponding to panels (c,f,i). Parameters are: $\bar{\textsl{g}}_{R}=2$, $\bar{\gamma}_{C}=3$, $\bar{\gamma}_{R}=15$, and (a-c) $\bar{R}=0.25$; (d-f) $\bar{R}\!=\!0.5$; (g-i) $\bar{R}=1$. }
\end{figure}

Quasi-1D solitons in the  open-dissipative polariton condensate behave in a different way. As demonstrated in Sec.~\ref{sec:DSDynamic}, dark solitons are fundamentally non-stationary excitations even without a transverse modulation,  and therefore the theory of transverse instability developed in Refs.~\cite{KuznetsovTuritsyn_1988, Opt.Lett._1993, Opt.Commun._1993, Europhys.Lett._1995, PhysRevE.51.5016_1995, Phys.Rep._2000, Tikhonenko_1996, PhysRevLett.76.2262_1996, PhysRevA.54.870_1996, PhysRevLett.86.2926_2001, PhysRevA.62.053606_2000, PhysRevA.65.043612_2002, Analogies_2004, PhysRevA.82.023621_2010, KamchatnovPitaevskii_ConvectiveInstability_2008, Kamchatnov_ConvectiveInstability_2011, MironovSmirnov_2011, HoeferIlan_2012} is not strictly applicable. Nevertheless, it is intuitively clear that transverse instability may occur and dramatically influence the process of relaxation of solitonic structures described in Sec.~\ref{sec:DSDynamic}. Indeed, the transverse instability of dark soliton stripes is clearly seen in numerical simulation of the polariton condensate dynamics in the framework of Eqs.~(\ref{eq:psieq}) and~(\ref{eq:mReq}). The results of numerical simulations for a weak pump, $\left(P_{0}\bigl/P_{th}\bigr.\!-\!1\!\ll\!1\right)$, are presented in Fig. \ref{fig:TrInstability_WeakPump_1}, where we plot the condensate density $n_C(\vec{r},t)$ [panels (a-i)] and phase ${\rm arg}(\psi)$ [panels (j-l)] at different moments in time. The reservoir density $m_R(x,y)$ (not shown)  trivially follows that of the condensate polaritons with density peaks corresponding to density dips in the condensate. 
  
At $t=0$, the wave function $\psi\!\left(\vec{r},t=0\right)$ was taken in the form~(\ref{eq:NumInitDistribution}), whereas velocity ${v_{s}}_{0}$ and parameters $\delta {v_{s}}_{0}$ and $\sigma_{y}$ were the same as in the case corresponding to Fig.~\ref{fig:NLS_1}. The behaviour of the polariton condensate was simulated for three different values of the dimensionless scattering rate $\bar{R}\!=\!0.25,{\,}0.5,{\,}1$. According to~(\ref{eq:tau_WeakPump}), for $\bar{R}\!=\!0.25$ the characteristic relaxation time $\tau_{1}$ for 1D dark soliton is equal to $30$, while for $\bar{R}\!=\!0.5$ it is two times less, and for $\bar{R}=1$ it is four times less. As seen in Fig.~\ref{fig:TrInstability_WeakPump_1}(a-c,j), a vortex pair is formed from the initial distribution, as a consequence of transverse instability development, at remarkably lengthy period of time ($t\!\approx\!50$), compared to the depicted in Fig.~\ref{fig:NLS_1} situation, where $t \leq 27.5$. After that, no new vortices form, the condensate excitations smoothly relax to the homogeneous state, while the vortex and antivortex  approach each other, annihilate and turn into a fading {\em vortex-free} localized structure. Growth of the stimulated scattering rate, $\bar{R}$, leads to inhibition of the transverse instability, whereby no vortex-antivortex pairs form, and slowly developing instability results in formation of vortex-free structures  [Fig. \ref{fig:TrInstability_WeakPump_1} (d-f,k)]. Finally,  if the lifetime of the solitonic structure, $\tau_1$, becomes comparable to the characteristic time scale of the linear stage of the transverse instability $1/\Gamma_{max}$, a dark soliton remains quasi-one-dimensional for all the time of its existence, the transverse instability becomes insignificant, and a solitonic stripe merges with the background without breaking up into two-dimensional localized structures. This regime is illustrated in  Figs. \ref{fig:TrInstability_WeakPump_1} (g-i,l), where $\tau_1=7.5$. For a condensate with the polariton decay rate $\gamma_c=10$ $ps$, this corresponds to $t=225$ $ps$.

\begin{figure}[t]
\includegraphics[width=8.5cm]{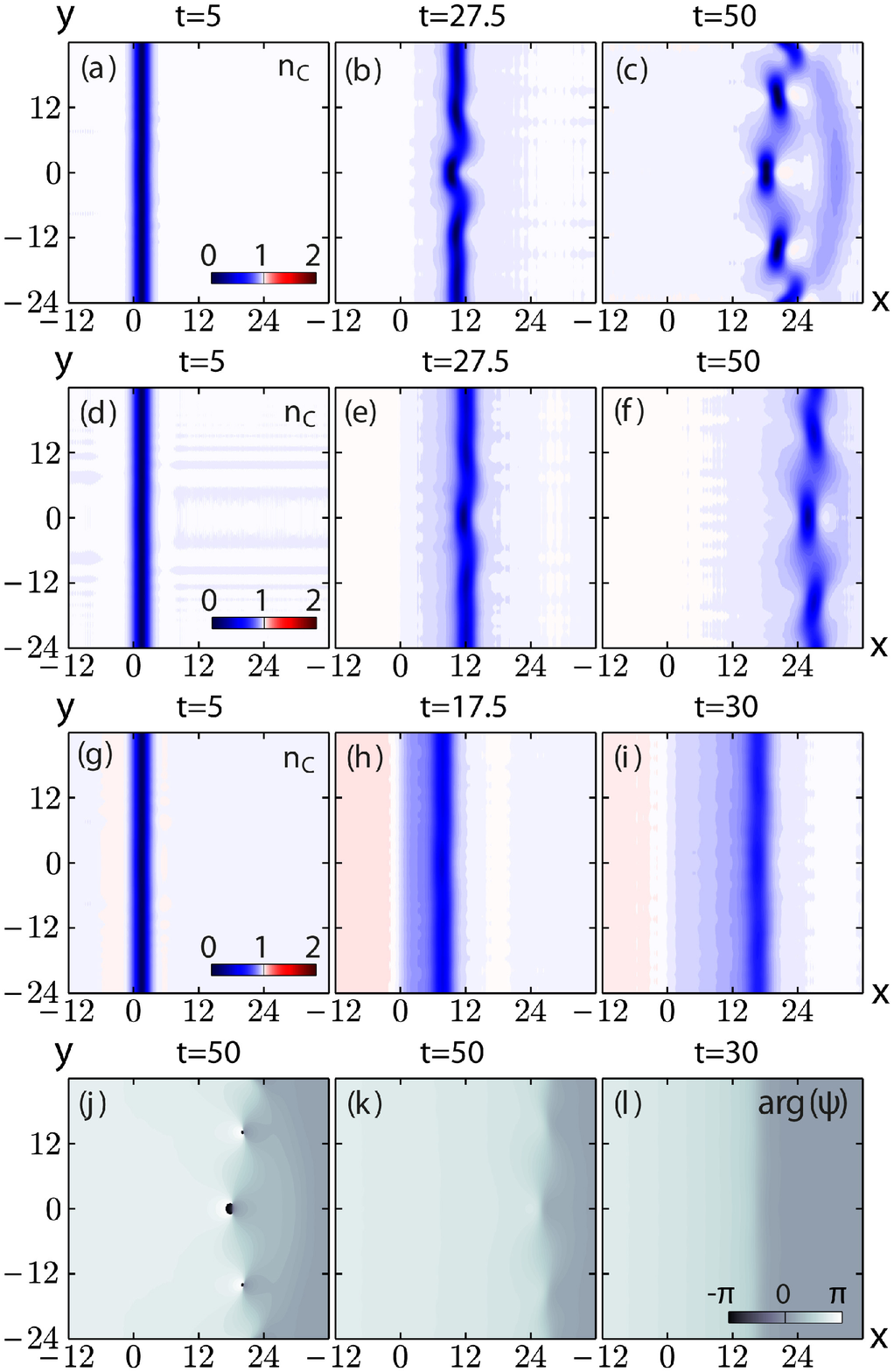}
\caption{\label{fig:TrInstability_StrongPump_1} Transverse instability development of a dark soliton with the initial velocity ${v_{s}}_{0}=0.3$ for strong pumping. The initial condensate wave function $\psi\!\left(\vec{r},t\right)$ is the same as that in Fig.~\ref{fig:NLS_1}, and $m_{R}\left(\vec{r},0\right)=0$. Shown are: (a-i) condensate density $n_C$, and (j-l) phase ${\rm arg}(\psi)$ corresponding to panels (c,f,i). Parameters are $\bar{\textsl{g}}_{R}=2$, $\bar{\gamma}_{R}=0.07$, $\bar{R}=0.08$, and (a-c) $\bar{\gamma}_{C}=0.0075$;  (d-f) $\bar{\gamma}_{C}=0.015$; (g-i) $\bar{\gamma}_{C}=0.03$.}
\end{figure}

In the case of strong pumping $\left(P_{0}/P_{th}\gg1\right)$ illustrated in Fig.~\ref{fig:TrInstability_StrongPump_1}, structures similar to those depicted in Fig.~\ref{fig:TrInstability_WeakPump_1} are formed. However, for the same initial conditions as in the case of weak pumping, we clearly observe the consequence of nonlocal coupling between the condensate $n_{C}\!\left(\vec{r},t\right)$ and reservoir density  $m_{R}\!\left(\vec{r},t\right)$. Indeed, the instability develops much slower  than in the case modelled by the NLS equation~(\ref{eq:NLS}) and leads to formation of several vortex pairs and two-dimensional vortex-free solitons which then disappear [Fig.~\ref{fig:TrInstability_StrongPump_1}(a-c,j)]. In this regime the soliton lifetime, $\tau_3$ is controlled by the renormalized decay rate, $\bar{\gamma}_{C}$. For larger decay rates (three times longer lifetime) the transverse instability development slows down and a vortex pair is no longer formed but three two-dimensional vortex-free solitons appear [Fig.~\ref{fig:TrInstability_StrongPump_1}(d-f,k)]. Further growth in the decay rate (and four-fold increase in $\tau_3$) fully inhibits the transverse instability, and the dark stripe retains its quasi-1D nature before merging with the background [Fig.~\ref{fig:TrInstability_StrongPump_1}(g-i,l)]. In order for a vortex pair to be formed, parameters of the condensate should be chosen so that $\tau_{3}>\tau_{1}$, which  indicates that, compared to the weak pumping case, it is more difficult to realize conditions for vortex-pair formation via the transverse instability development.

In addition, strong pumping leads to much shorter time scales of the soliton dynamics overall. For $\gamma^{-1}_c=10$ ${\rm ps}$, the physical periods of time for relaxation and instability development depicted in Fig.~\ref{fig:TrInstability_StrongPump_1} are $t=3.75$ ${\rm ps}$ (a-c), $t=7.5$ ${\rm ps}$ (d-f), and  $t=9$ ${\rm ps}$ (g-i), respectively.

From the above numerical analysis,  we conclude that vortex pairs in a polariton condensate may form due to development of the transverse instability of quasi-1D dark stripes, but only under very specific conditions and for a short period of time.
\section{Conclusions}\label{sec:Conclusion}

We have shown, within the framework of an open-dissipative mean-field model, that a homogeneous exciton-polariton condensate with an incoherent pump can support spatially localized solitonic excitations with the spatial and phase structure similar to that of one-dimensional dark solitons in conservative systems. These nonlinear excitations have a finite lifetime which is determined by the parameters of the system, such as polariton decay rates, stimulated scattering rate, and strength of polariton-polariton interactions. In the process of evolution, the dark solitonic structures grow in width and lose contrast eventually blending with the homogeneous background. The soliton relaxation depends on parameters of the condensate and proceeds differently in slightly and highly above-threshold pumping regimes. The characteristic soliton lifetimes, given by compact analytical formulas, could potentially assist experimental verification of our theory.

Finally, the scenario of transverse instability development in exciton-polariton condensate is shown to be different from that in atomic condensates governed by Gross-Pitaevskii (or NLS) mean-field model. For the latter, a nonlinear stage of transverse instability most frequently ends up with a dark soliton decaying into vortex-antivortex pairs. By contrast, in exciton-polariton condensate several rather strict conditions should be satisfied for vortex-pair formation via instability development. In the majority of regimes, a soliton breaks up into vortex-free two-dimensional localized structures, which disappear rather fast. The fact that the dark stripe in a two-dimensional condensate may relax by losing its contrast before the transverse instability takes place, could inhibit potential experimental observation of dark soliton decay into vortex-antivortex pairs in an incoherently formed exciton-polariton condensate.

Although our analysis is performed for a homogeneous condensate, it is applicable to realistic quasi-homogeneous condensate created by a broad beam {\em cw} excitation with a "top-hat" intensity profile.

\section{Acknowledgments}

This work was supported by the Australian Research Council through the Discovery and Future Fellowships programs. 
\bibliography{DarkSolitons}
\end{document}